\documentclass[sigconf]{acmart}

\AtBeginDocument{%
  \providecommand\BibTeX{{%
    \normalfont B\kern-0.5em{\scshape i\kern-0.25em b}\kern-0.8em\TeX}}}

\setcopyright{acmcopyright}
\copyrightyear{2022}
\acmYear{2022}
\acmDOI{10.1145/1122445.1122456}

\acmConference[ICSE SEIS '22]{ICSE SEIS '22: International Conference on Software Engineering, Software Engineering in Society}{May 21--29, 2022}{Pittsburgh, PA, USA}
\acmBooktitle{ICSE SEIS '22: International Conference on Software Engineering, Software Engineering in Society,
  May 21--29, 2022, Pittsburgh, PA, USA}
\acmPrice{15.00}
\acmISBN{978-1-4503-XXXX-X/21/06}

\usepackage{algorithmic}
\usepackage{graphicx}
\usepackage{textcomp}
\usepackage{xcolor}
\usepackage{comment}
\usepackage{fontawesome}
\usepackage{csquotes}
\usepackage{relsize,etoolbox}
\usepackage[rightmargin=0.5pt,leftmargin=0.5pt]{quoting}
\usepackage{booktabs} 

\usepackage{xspace}
\usepackage[normalem]{ulem}
\useunder{\uline}{\ul}{}
\usepackage{multirow}
\usepackage{lscape}
\usepackage{rotating}
\usepackage{supertabular}
\usepackage{longtable}
\usepackage{enumerate}
\usepackage{colortbl}
\usepackage{subcaption}
\usepackage{hhline}
\usepackage{makecell}
\usepackage{caption}
\usepackage{hyperref}
\usepackage{enumitem}
\usepackage{tcolorbox}
\usepackage{tabularx}
\usepackage{lipsum}
\usepackage{balance}
\usepackage{blindtext,graphicx}
\usepackage[absolute]{textpos}
\newcommand{\HK}[1]{{\color{red}{#1}}}

\newcommand{\todo}[1]{}
\renewcommand{\todo}[1]{{\color{red} TODO: {#1}}}

\def\BibTeX{{\rm B\kern-.05em{\sc i\kern-.025em b}\kern-.08em
    T\kern-.1667em\lower.7ex\hbox{E}\kern-.125emX}}

\begin{document}
%

\HK{\title{How are Diverse End-user Human-centric Issues Discussed on GitHub?}}



%
\author{Hourieh Khalajzadeh, Mojtaba Shahin, Humphrey O. Obie, John Grundy}
\affiliation{%
  \institution{HumaniSE Lab, Faculty of Information Technology, Monash University, Melbourne, Australia}}
\email{{hourieh.khalajzadeh, mojtaba.shahin, humphrey.obie, john.grundy}@monash.edu}



\begin{abstract}
Many software systems fail to meet the needs of the diverse end-users in society and
are prone to pose problems, such as accessibility and usability issues.
Some of these problems (partially) stem from the failure to consider the characteristics, limitations, and abilities of diverse end-users during software development. We refer to this class of problems as \textit{human-centric issues}. Despite their importance, there is a limited understanding of the types of human-centric issues encountered by developers. In-depth knowledge of these human-centric issues is needed to design software systems that better meet their diverse end-users' needs.
This paper aims to provide insights for the software development and research communities on which human-centric issues are a topic of discussion for developers on GitHub.
We conducted an empirical study by extracting and manually analysing 1,691 issue comments from 12 diverse projects, ranging from small to large-scale projects, including projects designed for challenged end-users, e.g., visually impaired and dyslexic users. 
Our analysis shows that eight categories of human-centric issues are discussed by developers. These include  Inclusiveness, Privacy \& Security, Compatibility, Location \& Language, Preference, Satisfaction, Emotional Aspects, and Accessibility.  
Guided by our findings, we highlight some implications and possible future paths to further understand and incorporate human-centric issues in software development to be able to design software that meets the needs of diverse end users in society.

\end{abstract}



\maketitle


%

\section*{Lay Abstract}

Many software systems fail to take into account diverse end user differences, such as age, gender, culture, language, physical and mental challenges, emotions, personality, and so on.  This means for many users the software is difficult if not impossible to use, unengaging, disrespectful, increases the digital divide, excludes many -- often vulnerable -- members of society, and may even be unsafe or dangerous. GitHub is a very popular software platform used by software developers. We looked at 
several diverse online software 
projects and the discussions developers have 
about what we call these "human-centric issues" in software. We learned that some issues are quite often discussed, 
however, many diverse end user characteristics are not well understood and many not often discussed by developers, suggesting they are not sufficiently well thought about during software development. We make some recommendations for software engineers to help them better consider and take account of many of their software user differences during development. This includes 
taking into account these important issues; for some projects some end user differences are more important than others depending on the target users; users need better ways of reporting human-centric defects and developers need better ways of addressing human-centric issues for software; and developer training to consider a variety of human-centric issues needs improving.

\section{Introduction} \label{sec:Introduction}
Software systems aim to deliver efficient and satisfactory solutions to fulfill the expectations of a wide range of diverse end users in society. However, complex software systems are prone to security and data breaches, massive cost overruns and project slippage, hard-to-deploy, hard-to-maintain, and even dangerous solutions and hard-to-use software \cite{grundy2020towards}. Many of these problems can be traced back to a lack of understanding and addressing of \textbf{\textit{human-centric issues}} during the software engineering process \cite{hartzel2003self,miller2015emotion,stock2008evaluation,wirtz2009age}. We define \textbf{\textit{human-centric issues}} as ``\textit{the problems that diverse users face when using a software system, due to the lack of (proper) consideration of their specific characteristics, limitations, and abilities}". These characteristics include differing personalities, technical proficiency, emotional reactions to software systems, socio-economic status, gender, age, culture, preferences, working environment, and language. Software users also access software from different locations through different devices and platforms, with some only being able to afford limited options. 


To be able to better design software that meets the diverse needs of end-users in society, such characteristics need appropriate consideration in all aspects of the software by developers. 
Many software solutions are developed by professionals who are not aware of, have not experienced, or do not understand and effectively communicate the implications of differing human-centric issues of their users. For example, the underlying reason for developing apps with poor accessibility issues has been shown to be a lack of awareness and training about accessibility and its importance among developers \cite{Alshayban:2020}. 
When handling human-centric issues in software design, developers need to be aware and carefully consider the characteristics, limitations, and abilities of the end-users~\cite{2007Kulyk}.
Lack of consideration of these human-centric characteristics leads to the software -- which should primarily be designed and built to solve human needs -- not meeting the end-users' expectations and causing frustration, accessibility, and usability issues \cite{curumsing2019emotion,rafael2013cooperative,yusop2016reporting}.
Some studies have previously explored particular human-centric issues (e.g., accessibility), developer's issues and characteristics (e.g., emotions) \cite{Murgia:2014, Alshayban:2020, Rauf:2020}, or specific aspect of software development, such as UI/UX \cite{maguire2013using,ovad2015teaching}. However, there is still very limited evidence-based knowledge about how different types of end-user human-centric issues are discussed and addressed during software development. This work aims to understand: 1) whether developers discuss these human-centric issues, and 2) provide an in-depth and comprehensive understanding of different types of human-centric issues developers discuss and how they discuss them during the software development. 

Developers' discussions can be a major factor in deciding how a system evolves, suggesting that the discussions include information beyond how a system works \cite{brunet2014developers,tsay2014let}. Online software repositories, e.g., GitHub, attract a lot of discussions between developers on a variety of different topics. These repositories provide developers with perspectives on the issues they face during the software development process and how they react to them. They play a significant role in improving the capabilities of software developers/users and accelerating software development \cite{mo2015tbil}. Analysing the comments that developers leave in response to the issues might reveal the consideration of diverse end-users' human-centric issues from the viewpoint of developers.

To this end, we manually analysed 1,691 issue comments collected from 12 GitHub repositories. We considered a diverse range of applications, including apps designed for vulnerable users (e.g., visually impaired and dyslexic users), large scale end-user based project (Firefox), and software designed for unforeseeable situations (COVID 19 apps). Our analysis revealed that human-centric issues can be classified into eight categories: \textbf{Inclusiveness}, \textbf{Privacy \& Security}, \textbf{Compatibility}, \textbf{Location \& Language}, \textbf{Preference}, \textbf{Satisfaction}, \textbf{Emotional Aspects}, and \textbf{Accessibility}. Based on our findings, Privacy, Preference and Satisfaction are more often discussed by developers, while developers seem to discuss less Emotional aspects and Accessibility related issues. COVID 19 apps (COVIDSafe Australia and Corona-Warn-App Germany) include more human-centric discussions (Privacy, Preference and Satisfaction), while general purpose and health apps have fewer human-centric discussions. The main contributions of this work include: 
\begin{itemize}
\item Manually analysing a relatively large number of issue comments from 12 GitHub repositories, and identifying eight categories of human-centric issues;
\item Providing some implications and possible future research directions to better manage human-centric issues in software development, aiming to meet the needs of society; and
\item Building and publicly releasing a replication package to enable researchers and practitioners to access all collected data and replicate and validate our study \cite{anonymous_2021_4739069}.
\end{itemize}

The rest of the paper is structured as follows. Section \ref{sec:background/motivation} provides the background and motivation of this study. Section \ref{sec:researchmethod} presents our research methodology. Section \ref{sec:findings} presents the study results. Section \ref{sec:Discussion} reflects on the key findings. Section \ref{sec:ThreatsValidity} lists the possible threats to validity of our study. Section \ref{sec:RelatedWork} reviews key related work. Finally, Section \ref{sec:Conclusion} draws conclusions and proposes avenues for future work.

\section{Motivation}
\label{sec:background/motivation}

\subsection{Motivating Example}
\begin{figure}[b!]
\centering
\includegraphics[width=0.9\linewidth]{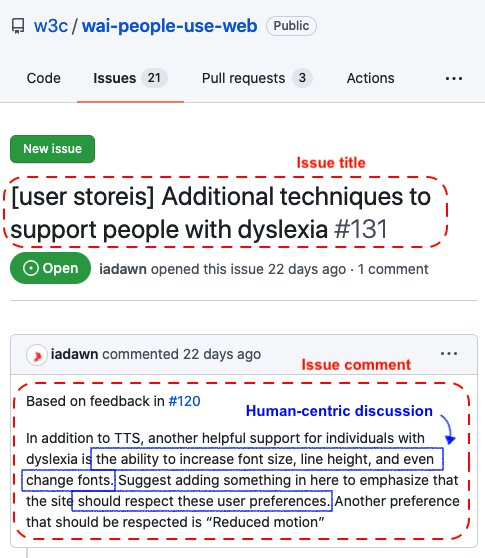}
\caption{An example human-centric issue from GitHub}
\label{fig:Motivation}
\end{figure}
Imagine a dyslexic person who wants to access a website to get some information on their diet. This user might have specific requirements to be able to access the website content. As one of the most popular software repositories, the issue tracker in GitHub provides an option for end-users and developers to report issues and provide feedback on a software system (e.g., a diet website) hosted on GitHub.

A discussion in the issue tracker initiates with a title (\emph{issue title}), followed by subsequent posts (\emph{issue comments}) from reporters and contributors, including project maintainers, developers, users, or the reporter itself. 
Figure \ref{fig:Motivation} shows such an issue in the GitHub issue tracking system made by a collaborator to discuss the dyslexic user's preferences. This is followed by a comment from another collaborator listing some other barriers and asking whether this is true: \textit{``I'm not sure I agree with it''}. The reason for such disagreement is probably that the developer is not fully aware of the needs and preferences of the user. However, discussing such issues can help developers be aware of such challenges and consider such issues when designing software. 
This example shows the importance of paying attention to and discussing issues related to human aspects (i.e., we refer to such issues in this paper as \emph{human-centric issues}) in the uptake of the software. 

Awareness of and discussing human-centric issues may lead to designing more inclusive software for all members of society. Negligence of these issues can exclude diverse users of society from accessing the software. Such issues are not limited to the users with special needs. As another example, if an app has compatibility issues, it excludes a group of users with a specific device or software from using it. If an app does not provide different languages, it excludes the users who do not understand the provided languages. 
Therefore, there is a need for better understanding and supporting human-centric issues to be able to design software to meet the needs of the whole society.  In this work, we are interested to understand how such human-centric issues are discussed in GitHub, and believe promoting awareness of such issues helps better accounting for them, and therefore designing more inclusive software. 

\subsection{Human Aspects}
\label{sec:humanaspects}
Understanding such end-user human-centric issues, or human aspects, plays an essential role in designing software that meets the requirements of diverse users of the society. Such human aspects include age, gender, culture, language and location, digital literacy, physical and mental impairments, and also emotional impacts of the software on users due to their diverse personalities and preferences \cite{grundy2020humanise}. Lack of consideration of diverse users' \textbf{preferences} and \textbf{satisfaction}, leads to human-centric issues when using the software. Different \textbf{age} groups have different expectations, challenges, and reactions to the same software \cite{hussain2017ux}. 
\textbf{Cultural} differences significantly influence the uptake of the software. Users speak different \textbf{languages} and access the software from various \textbf{locations} all around the world. \textbf{Gender} bias in software applications, such as smart living technologies \cite{perez2019invisible,strengers2020smart}, reflect the importance of taking gender-related issues into account when designing a software system.  \textbf{Physical and mental impairments} of end-users impact the ways they are able to access the software. Different users have various \textbf{emotional} reactions to the software and such emotional impacts can influence the uptake of applications \cite{miller2015emotion}. Therefore, to be able to design software that meets the requirements of the whole society, such human-centric aspects need to be well understood, discussed, and incorporated in the software development. Taking human-centric issues into account can have a huge impact on diversity, inclusion, belonging, and representation of vulnerable groups of the users in society.


\begin{figure*}[t]
\centering
\includegraphics[width=0.95\linewidth]{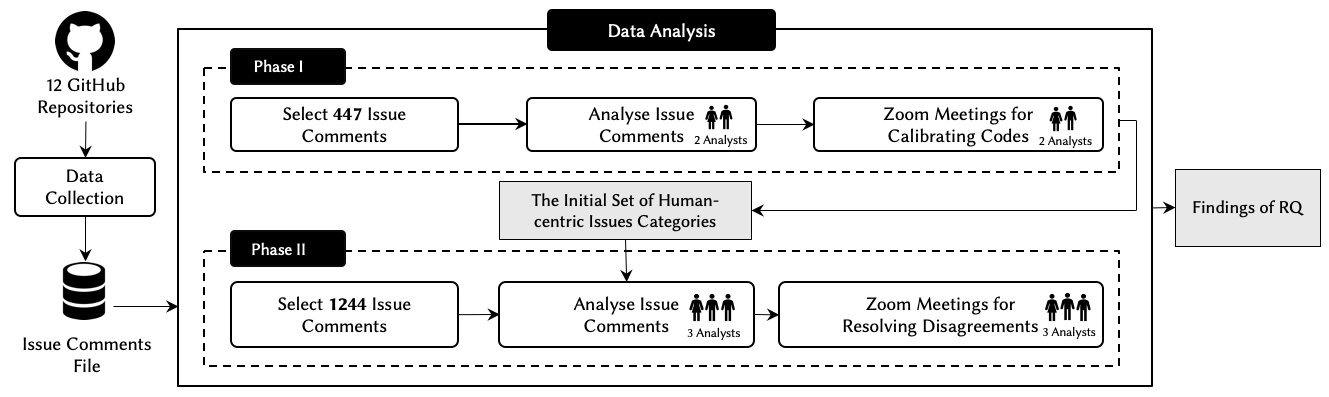}
\caption{An overview of our research method} 
\label{fig:researchmethod}
\end{figure*}

\section{Research Method}
\label{sec:researchmethod}


Our study is motivated by the need to help practitioners and researchers be more aware of different types of diverse end user human-centric issues occurring in software project's lifecycle and identify possible areas for improvement and investment in addressing these. This would ultimately help in the design of software that better meets diverse end-users needs.
Hence, we formulated the following research question:

\begin{tcolorbox}
\textit{\textbf{RQ. What end-user human-centric issues do developers typically discuss in their GitHub repositories?}}
\end{tcolorbox}


To answer this research question, we conducted an empirical study on a subset of issue comments collected from 12 GitHub repositories. Figure \ref{fig:researchmethod} presents an overview of our research method. We detail our research method in this section.


\subsection{Projects Selection}\label{sec:projectselection}
We selected 12 projects hosted on GitHub. Table \ref{tbl:projectslist} shows details of these projects. We deliberately focused on end-user-based projects with different sizes, domains, and types of users to increase the chance of identifying human-centric issues. 
These projects can be generally categorised in four groups:

\textit{\textbf{Apps with millions of users}}. In this group, we selected Firefox for iOS, one of the most popular open-source mobile web browsers hosted on GitHub with millions of users. Firefox has been extensively studied in software research from different technical aspects such as security (e.g., \cite{zaman2011security}) and release engineering (e.g., \cite{khomh2012faster}). However, it has not yet been studied from human aspects. As Firefox attracts large number of users with different characteristics (e.g., different ages, levels of education), we expected this may increase the chance of discussions on human-centric issues among developers.
 
\textit{\textbf{COVID-19 contact tracing apps}}. Governments have been developing COVID-19 tracing apps as an effective approach to control the COVID-19 pandemic. Despite their effectiveness, this class of apps is associated with numerous social- and human-related issues such as privacy concerns \cite{cho2020contact} and ethical issues \cite{parker2020ethics}. Among the existing COVID tracing apps, we chose the COVIDSafe app developed by the Australian government and Corona-Warn-App developed in Germany since both apps have GitHub repositories and Android and iOS versions.
 
\textit{\textbf{Healthcare apps/tools}}. We focused on healthcare apps/tools as this type of apps may pose significant risks to patients and healthcare professionals \cite{lewis2014mhealth}. The possible risks are enormous and range from loss of privacy and reputation to loss of life. 
We chose HealthChecks as it is the most popular healthcare-related GitHub repository (it has the highest number of stars in healthcare-related repositories on GitHub).
  
\textit{\textbf{Apps for vulnerable users}}. We expected developers would talk more about human-centric issues (e.g., accessibility) when developing apps for vulnerable users, especially those needing visual features (e.g., visually challenged people). We searched through the Google app store to find the popular apps for vulnerable users. We looked for terms ``dyslexia", ``visually impaired", ``blind people" and found 58 related apps with at least one app review on the app store. We then looked for the ones with a GitHub repository, i.e., 35 apps and selected the ones with at least one open issue. This gave us six projects as \textit{NavCog} and \textit{Corsaire}, designed for visually impaired users, and \textit{Opendyslexic-chrome}, \textit{eBookDys}, \textit{Opendyslexic-firefox}, and \textit{predict4all} target dyslexic users.     


\subsection{Data Collection}\label{sec:datacollction}

As discussed, developer discussions (issue comments) in issue tracking systems include a wide range of rich and detailed information about user needs, essential design decisions, the rationale behind decisions, bugs, faults, etc. Hence, we leveraged issue discussions from our 12 projects as the potential source to identify human-centric issues discussed by developers. In total, the 12 projects had 12,088 
issue comments that we extracted using GitHub v3 API. 

\begin{table*}[]
\centering
\caption{List of projects studied in this paper}
\vspace{-3mm}
\label{tbl:projectslist}
{\scriptsize
\renewcommand{\arraystretch}{1.5}
\begin{tabular}{|l|l|l|c|c|c|c|c|}
\hline
\textbf{Project Name}                                    & \textbf{Repository Name}                                    & \textbf{URL}                                & 
\textbf{\begin{tabular}[c]{@{}l@{}}\# Issue Comments\end{tabular}}          & \textbf{\# Contributors}       & \textbf{\# Watch}         & \textbf{\# Star}            & \textbf{\# Releases}    \\ \hline \hline
COVIDSafe-ios & AU-COVIDSafe/mobile-ios          & https://bit.ly/3sosjuq 
& 34                &    2            &    57     &     279   &      N/A     \\ \hline
COVIDSafe-android &AU-COVIDSafe/mobile-android      & https://bit.ly/39zZouF 
& 46                &   2             &     60    &   364     &      N/A     \\ \hline
Firefox & mozilla-mobile/firefox-ios       & https://bit.ly/2LszT6O 
& 8228                & 190               &     483    &    9.9k    &     121      \\ \hline
Cwa-app-ios & corona-warn-app/cwa-app-ios      & https://bit.ly/3oNJao6 
& 2010                &  72              &     88    &    1.5k    &      145     \\ \hline
Cwa-app-android & corona-warn-app/cwa-app-android  & https://bit.ly/3oN5M8v 
& 340                &   77             &    120     &     2.2k   &    112       \\ \hline
HealthChecks & healthchecks/healthchecks        & https://bit.ly/3srT1Cs 
& 1227                &    55            &    90     &  3.4k      &     24      \\ \hline
NavCog & hulop/NavCogIOS                  & https://bit.ly/3nIYyAP 
& 99                & 5              & 13      & 10      & 5         \\ \hline
Corsaire & snigle/corsaire                  & https://bit.ly/3oKjoBe 
& 19                & 1              & 3       & 5      & N/A       \\ \hline
OpenDyslexic-chrome & OpenDyslexic/opendyslexic-chrome & https://bit.ly/2LNdDEm 
& 61                & 8              & 4       & 35     & 19        \\ \hline
eBookDys & garconvacher/eBookDys            & https://bit.ly/2XJWZIm 
& 3                & N/A              & 3       & 6      & N/A       \\ \hline
OpenDyslexic-firefox & OpenDyslexic/firefox-extension   & https://bit.ly/3oNw6z8 
& 19                & N/A            & 2       & 6      & N/A       \\ \hline
Predict4all & mthebaud/predict4all             & https://bit.ly/39w4xEh 
& 2                & N/A            & 2       & 1      & 2       \\ \hline
\end{tabular}
}
\end{table*}
\subsection{Data Analysis}\label{sec:dataanalysis}
We conducted our data analysis in two phases:
\subsubsection{Phase I} In the first step, we randomly selected 244 issue comments from 
\textit{HealthChecks} (25 issue comments), \textit{Firefox} (178 issue comments), \textit{Cwa-app-android} (12 issue comments), and \textit{Cwa-app-ios} (29 issue comments). As \textit{NavCog}, \textit{Corsaire}, \textit{Opendyslexic-chrome}, \textit{eBookDys}, \textit{Opendyslexic-firefox}, and \textit{predict4all} had a relatively small number of issue comments, we chose all issue comments (203) from these projects. The first two authors (i.e., analysts) independently inspected and classified the 447 (244+203) issue comments by following the open coding technique \cite{glaser1968discovery}. At the end, each analyst developed a list of the human-centric issues he/she found in the 447 issue comments. Then, the analysts held several Zoom meetings to check similarities and differences between their analysis and labelling and calibrate the identified codes. These meetings led to constructing an initial set of categories of human-centric issues: Inclusiveness, Privacy \& Security, Compatibility, Location \& Language, Preference, Satisfaction, Emotional Aspects, and Accessibility. The analysts also jointly provided a precise definition for each of the categories.
\subsubsection{Phase II}
In the second phase, we adopted power statistics \cite{kadam2010sample} to calculate a proper sample size of issue comments in each project. At the 95\% confidence level, we set a 5\% margin of error and randomly selected the following number of issue comments (excluding the 244 samples selected for the first phase) from each project: 
\textit{HealthChecks} (293 issue comments), \textit{Firefox} (367 issue comments), \textit{Cwa-app-ios} (323 issue comments), \textit{Cwa-app-android} (181 issue comments), \textit{COVIDSafe-android} (46 issue comments), and \textit{COVIDSafe-ios} (34 issue comments). 
Given the apps designed for visually impaired and dyslexic users had a limited number of issue comments, and were analysed in the first phase, we did not further consider them in the second phase. The first three authors (i.e., analysts) analysed these 1,244 issue comments. Each of the analysts manually analysed 830 individual issue comments. In other words, each issue comment was analysed and labeled by two analysts. Based on the initial categories that emerged from Phase I, we created a spreadsheet and shared it with three analysts. The analysts were asked to indicate whether an issue comment included at least one human-centric issue. If so, they had to specify which of the initial categories of human-centric issues the given issue comment belonged to and put ``1'' in the corresponding columns in the spreadsheet. Comments could be coded with more than one issue. Although the analysts had the freedom to capture and add any new human-centric issues category that they felt did not belong to Phase I's initial set of categories, no new human-centric issues categories were found. While this does not follow the idea of open coding, this decision was made for two reasons. First, it avoided developing a potentially very large number of possible human-centric issue categories \cite{humbatova2020taxonomy}. Second, it supported the analysts to reach and use consistent labelling without introducing substantial bias \cite{humbatova2020taxonomy}. 

Finally, the three analysts held two 3-hour Zoom meetings to compare their labelling results and resolve possible disagreements. The majority of disagreements were resolved through discussions between the two assigned analysts by providing the reason behind their choices explicitly. If the two analysts could not reach an agreement, the third analyst was asked to read and label the conflicting issue comment. Then, we voted to resolve the disagreement.

\section{Findings}
\label{sec:findings}
Based on our analysis of 447 issue comments in the first phase of the study and 1,244 further issue comments in our second phase, we determined eight broad end-user human-centric categories in the GitHub discussions: \textbf{Inclusiveness}, \textbf{Privacy \& Security}, \textbf{Compatibility}, \textbf{Location \& Language}, \textbf{Preference}, \textbf{Satisfaction}, \textbf{Emotional Aspects}, and \textbf{Accessibility}. In this section, we provide definitions of these categories, and a summary of their prevalence across different repositories.

\subsection{Human-Centric Issues Categories}

\subsubsection{\textbf{Inclusiveness}}
This category covers all the issue comments that discuss inclusion issues or exclusion of specific groups of users. If the discussion relates to discrimination toward a specific group of users, it falls into this category. It also includes issues related to the age and gender, and socio-economic status of the users. For example, one of the issues raised 
in the German COVID 19 app repositories is: 
\begin{quoting}
\noindent\faComment \hspace{0cm} \textit{``\textbf{Using a German App Store account is just not possible for many of us}, ... I am surprised we don't hear more in the media about this \textbf{large group of people being locked out of participating with the app}." }- (Cwa-app-ios)

\end{quoting}

This comment discusses an \textbf{Inclusiveness} related issue due to the location of user. Another example relates to the details not provided to the user causing a large majority of the users not aware of it, to be excluded from using the app:

\begin{quoting}
\noindent \faComment \hspace{0cm} \textit{``You need to start the app manually after each reboot. ... \textbf{nearly NO USER knows this.}
I think you \textbf{highly overestimate the percentage of people who even semi-regularly shutdown/reboot their phone.}" }- (Cwa-app-ios)

\end{quoting}

The inclusion of users of different ages is another human-centric issue discussed by developers: 

\begin{quoting}
\noindent \faComment \hspace{0cm} \textit{``It's important to \textbf{ensure kids are able to participate in society}." }- (COVIDSafe-ios)
\end{quoting} 
\subsubsection{\textbf{Privacy \& Security}}
This category covers all the issue comments related to privacy, security, data protection, reliability, and trust. Furthermore, we classified developer discussions concerning accessing the location and private data of a user into this category. Most of the privacy-related issues we found are related to accessing the location of the user, questioning why this is required, and whether asking for users' permission means their location is being tracked. There are also discussions emphasising that \textbf{ users' identities should not be revealed}. Another interesting topic in this category was the change of privacy in different versions of the app. 


\begin{quoting}
\noindent \faComment \hspace{0cm} \textit{``The app always required \textbf{location permission}. You would had previously given the app ``fine" location permission. However, as of v1.0.39 it now requires "coarse" location permission instead (but \textbf{doesn't use your existing ``fine" permission that it already has})." }- (COVIDSafe-android)

\end{quoting} 

The main concern was related to the apps accessing the location of the users. For example, there was a concern regarding having to enable location mode in order to get Android to locate the Bluetooth device, and the fact that if its disabled, every Android device will send the same Bluetooth-ID. One of the developers' responses to this concern was that:

\begin{quoting}
\noindent \faComment \hspace{0cm} \textit{``An app \textbf{does not get permission from the user to access location}, and then still tracks the user's location by BLE-scanning ... but \textbf{leads to the requirement to ask the user for location permission, even though the location isn't used within the app.}" }- (Cwa-app-ios)

\end{quoting} 

On the other hand, there is a discussion among developers in COVIDSafe repository on whether to allow people to use the app even without location, as long as they are warned that the app is not working, and ask them if they would want to turn it on or not. A developer reacted to this concern as follows: 

\begin{quoting}
\noindent \faComment \hspace{0cm} \textit{``The app is unable to get any Bluetooth permissions if the \textbf{location permission} is not enabled and \textbf{the app would not function at all due to Android policy.}" }- (COVIDSafe-android)

\end{quoting}

Another developer's response to this issue was: 

\begin{quoting}
\noindent \faComment \hspace{0cm} \textit{``In order for the app to be fully functional (i.e. able to detect other phones running COVIDSafe) then you \textbf{must have location enabled} and the location permission granted. If location is disabled, then the app will still be detectable by other phones (and they connect to you). As all exchanges are bidirectional, this means that you'll successfully encounter log the other phones (and they'll log you), but of course \textbf{this requires that everyone else has location enabled so that they can detect and connect to you.}" }- (COVIDSafe-android)
\end{quoting}


\subsubsection{\textbf{Compatibility}}
Any discussions around the compatibility of an app with different devices, operating systems, and platforms are included in this category. Compatibility issues are normally thought of as technical, not human-centric issues. However, a common reason for them occurring can be because of the users' socio-economic status, i.e., not having access to the latest phones, or the developers' ignorance, i.e., not taking all different platform choices into account. An example of developers not taking into account compatibility in earlier stages of developing software is:

\begin{quoting}
\noindent \faComment \hspace{0cm} \textit{``I found out that dp3-t uses the old bluetooth api ... which \textbf{is apparently not compatible with the Google/Apple protocol}."} - (Cwa-app-ios)

\end{quoting}

Compatibility-related issues may lead to some functionalities and features not accessible for a group of users. As an example, we found an issue comment discussing installing an updated version of Google Play Services would cause the app to crash, and the suggested solution was: \textit{``\textbf{having an outdated Play Services version}."} followed by asking whether \textit{``someone figured out whether the function calls that cause the crash can be disabled?"} - (Cwa-app-android)

Compatibility issues can exclude a specific group of users as these type of issues demotivate users to use the software. An interesting example relates to a trade-off between using the notification API and supporting the older iOS versions, that would exclude the users with older iOS versions: 

\begin{quoting}
\noindent \faComment \hspace{0cm} \textit{``If you will not use the new notification API, \textbf{then you can support older iOS version}, what is more important." }- (Cwa-app-ios)

\end{quoting}

This issue is further raised by another developer as:

\begin{quoting}
\noindent \faComment \hspace{0cm} \textit{``Currently a lot of people are angry at the government because \textbf{the warn app does not work with older Android versions.}" }- (Cwa-app-android)

\end{quoting}

An example of avoiding  compatibility with older versions in  the  first  implementation, raised by a developer is:

\begin{quoting}
\noindent \faComment \hspace{0cm} \textit{``I'd first \textbf{focus on supporting android 6.0+ here}. ... Adding legacy support to our main approach adds another layer of complexity which we should \textbf{avoid in our first implementation.} Our goal at the moment should be to at least be able to have full FLOSS version of the app." }- (Cwa-app-android)

\end{quoting}

Compatibility-related issues can force users to spend extra costs, if they can afford to, in case they need an app, as reflected vividly in the following issue comment:

\begin{quoting}
\noindent \faComment \hspace{0cm} \textit{``The app is just a front end for an API that was developed by Google. And \textbf{Google of course wants to sell more phones, so they only implemented it for newer Android versions, not for the old ones}." }- (Cwa-app-android)

\end{quoting}

A common solution suggested in discussions was to implement new APIs to make the app available for older phones as well, that would also allow to back-port the algorithm to older phones and therefore customize everything. 

\subsubsection{\textbf{Location \& Language}}
Any issues related to the physical location from where the user is accessing the software. These also include discussions about language or culture-related issues -- not always fully aligned with user location but often so. Based on our analysis, users' access may be limited if they are visiting a country and have no local phone number or App store account. For example, a comment related to an issue faced by a Luxembourger person living in Germany is:

\begin{quoting}
\noindent \faComment \hspace{0cm} \textit{``\textbf{the app being only available on the German app store whether it is iOS or Android should be resolved quickly.} ... \textbf{only those on the German app store can download it}. Clearly the politicians have dismissed to ask for an app that is available worldwide which would have made more sense. We're EU, Schengen, Open-Borders." }- (Cwa-app-ios)

\end{quoting}

Another example of a location-based issue that also reflects Inclusiveness related issues emphasises the need for the app to be translated into as many languages as possible, preferably all languages spoken in Germany: 

\begin{quoting}
\noindent \faComment \hspace{0cm} \textit{``... the idea of CWA is not to be a commercial app but rather a \textbf{social service the value of which increases the more people use it}. If this requires to support all iOS language then be it so..." }- (Cwa-app-ios)

\end{quoting}





\subsubsection{\textbf{Preferences}}
Any discussion related to the user's preferences from the point of view of the users or the developers viewpoint fall into this category. This relates to the features or functionalities that users prefer based on their specific human characteristics. Preference-related discussions include different aspects: (1) requesting new features (2) issues or requests to change an existing feature, such as the position of user interface elements (3) and privacy-related issues due to personal reasons. Preferences are sometimes discussed according to users' feedback received through app reviews or by developers from the users' perspective.
In some apps, 
developers often use the app themselves and discuss their usage experiences on GitHub. A developer discussing a feature according to the users' need is: 

\begin{quoting}
\noindent \faComment \hspace{0cm} \textit{``widgets are meant to be highly contextual. The items in a widget can change, \textbf{depending on whether it's relevant to users' need or not}. So, instead of disabling the icon, we should hide it when there's nothing to erase." }- (Firefox)

\end{quoting}

Another example is: 

\begin{quoting}
\noindent \faComment \hspace{0cm} \textit{``Doing the way you suggested the fade out animation on the table view cell would start at the same moment the user tapped the cell. Meaning \textbf{the user would see the cell animating to the deselected state.} I did the other way because I didn't want to do that since all other cells are always selected until the user get back to that tableView." }- (Firefox)

\end{quoting}




An example of a comment discussing a privacy-related issue that also concerns users' preferences is related to disallowing screenshots. It was discussed that \textit{``Disallowing screenshots in an app globally is \textbf{preventing users from} documenting their status." }- (Cwa-app-android) 
A user requested that screenshots should be permitted to allow them to visualise their health status (if they want this) and save it as Screenshot, unless it shows \textbf{sensitive data}". However, a developer has provided a comment that due to very strict time limitations, it is \textit{\textbf{``not always possible to find a satisfying compromise for all parties involved."} }- (Cwa-app-android) and therefore, they decided to prevent in-app screenshots for all screens for the version.



\subsubsection{\textbf{Satisfaction}}
Any discussions of the users' satisfaction, dissatisfaction, and pleasure falls into this category. This includes discussions around users' complaints, battery usage problems, and spam messages. An example of such issue raised by a developer is: 

\begin{quoting}
\noindent \faComment \hspace{0cm} \textit{``With this new approach every time we go between a whitelisted page to a non whitelisted page we'd have to load/unload lists. \textbf{Imagine if a user had whitelisted reddit. every time they tapped on an external link they'd load all the lists into the tab. every time they went back all the rules would be added back in."} }- (Firefox)

\end{quoting}

A satisfaction issue related to battery usage that causes dissatisfaction if someone has no possibility to recharge the phone for a long period of time is captured in this comment: 

\begin{quoting}
\noindent \faComment \hspace{0cm} \textit{``One can try to estimate what it means in absolute numbers. In my case the \textbf{battery drained} from 100\% to 15\% in 24h (85\%). Multiplying with 26\% relative usage of (Warn App + Covid 19 Exposure Logging) this translates to an absolute 22\% battery drain within 24 hours." }- (Cwa-app-ios)

\end{quoting}
Another satisfaction-related issue with COVIDSafe is that it interacts poorly with ColorOs battery optimisation features on Oppo phones. This is also a compatibility-related issue. Unless someone can find a way to permanently disable battery optimisation, nothing can prevent it from happening.


\subsubsection{\textbf{Emotional Aspects}}
This category includes the possible emotional impacts that the software can have on the users, including making the users confused, worried, scared and bored when using the app. As an example, a developer indicated how a design decision may frustrate the users:

\begin{quoting}
\noindent \faComment \hspace{0cm} \textit{``Merging everything into a single setting is simple and easy to understand but it could also \textbf{frustrate users} that ONLY want to protect the passwords and not the app." }- (Firefox)

\end{quoting}

Another example raised by a developer is that:

\begin{quoting}
\noindent \faComment \hspace{0cm} \textit{``Anyone using this service knows the \textbf{anxiety of not having absolute and at-a-glance insight into their operations."}} \\ - (HealthChecks)
\end{quoting}
 
\subsubsection{\textbf{Accessibility}}
This category covers issue comments discussing accessibility issues. Discussions about the users with physical and mental impairments also falls into this category. For example, an accessibility issue as a side effect of dark mode discussed by a developer, and advised to be left as-is is: 

\begin{quoting}
\noindent \faComment \hspace{0cm} \textit{``\textbf{Changing the color of the accessory view for a disabled row state would be non-standard then}! It just seems to us that the indicator is enabled if its colour is more contrast than the background. ... this is a side-effect of the dark mode ... \textbf{I think it should be left as-is}, and I'll update those bugs with an explanation." }- (Firefox)

\end{quoting}
Another technical accessibility related issues is: 

\begin{quoting}
\noindent \faComment \hspace{0cm} \textit{``P1 for \textbf{accessibility} because users will be extremely confused and might inadvertently activate controls they don't intend to activate." }- (Firefox)
\end{quoting}

\subsection{Human-Centric Issues in Different GitHub Repositories} \label{sec:categoriesProjects}
\begin{figure}[h]
\centering
\includegraphics[scale=0.55]{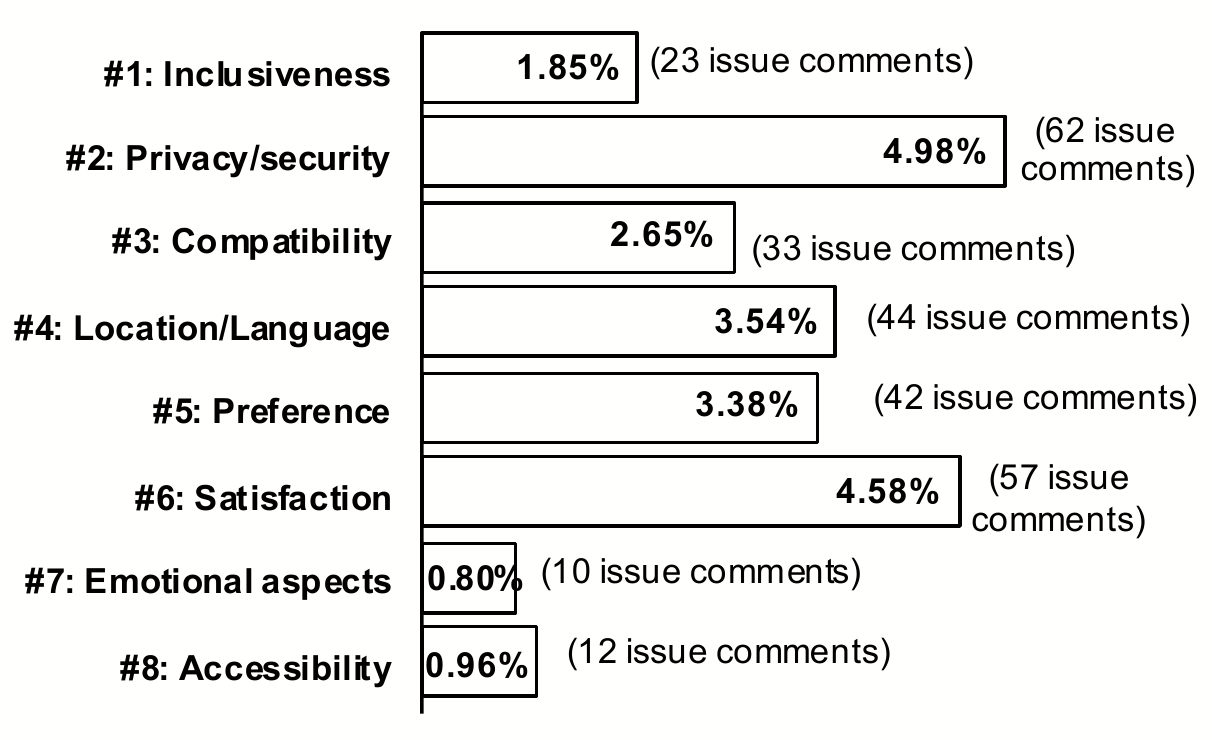}
\vspace{-3mm}
\caption{Number and percentage of human-centric issues out of 1,244 issue comments in Firefox, Cwa-app-ios, Cwa-app-android, HealthChecks, COVIDSafe-ios, COVIDSafe-android}
\label{fig:percentages}
\end{figure}

Of the 12 studied projects, six (apps designed for visually impaired and dyslexic users) were small and had a very limited number of issue comments. Hence, we do not compare and contrast the prevalence of human-centric issues of these projects in this section. Table \ref{tbl:Table2} provides detailed information on the human-centric issues categories in the issue comments of the six remaining projects (\textit{Firefox}, \textit{Cwa-app-ios}, \textit{Cwa-app-android}, \textit{HealthChecks}, 
\textit{COVIDSafe-ios}, \textit{COVIDSafe-android}).
Overall, 22.74\% of the comments (283 out of 1,244 issue comments studied in these six projects) discuss human-centric issues. How these human centric issues spread among different categories, is shown in Figure \ref{fig:percentages}. 

Among the 1,244 analysed issue comments, as shown in Figure \ref{fig:percentages}, privacy \& security (62 issue comments, 4.98\%) and satisfaction (57 issue comments, 4.58\%) are the main issues discussed by developers. Issues related to the location \& language (44 issue comments, 3.54\%), preference (42 issue comments, 3.38\%), compatibility (33 issue comments, 2.65\%) are the next ones. Inclusiveness-related issues are only discussed in 23 issue comments (1.85\%). Accessibility issues (12 issue comments, 0.96\%) and emotional effects (10 issue comments, 0.8\%) are found very rarely in the discussions. However, since we are not analysing how often the other issues, such as refactoring, technical debt, performance, and so on, are discussed, we can not comment on whether they are discussed to a limited extent or not. This is out of the scope of this paper, and we encourage future research on comparing the human-centric issues with other issues.

Comparing different projects, according to Table \ref{tbl:Table2}, COVID 19 apps (COVIDSafe Australia and Corona-Warn-App Germany) include more human-centric issues related discussions, whereas the other apps have very limited discussions of human-centric aspects. Privacy, preference, and satisfaction are the most frequent issues discussed in the COVID 19 apps. Compatibility and issues related to the location of the users, including language, are the next topics of discussion. Inclusiveness, as a result of compatibility and location, is discussed to an extent, while the accessibility and emotional impacts of the way the apps and their interfaces are designed are rarely discussed. Firefox (47 issue comments, 12.8\%), and HealthChecks (34 issue comments, 11.6\%)
, although designed for a large population of users, sometimes with health issues, do not include many human-centric issues discussions.

   
\begin{table*}[]
\centering
\caption{Number (\#) and percentage (\%) of human-centric issues in different projects}
\vspace{-3mm}
\label{tbl:Table2}
{\scriptsize
\renewcommand{\arraystretch}{2}
\begin{tabular}{|l|l|l|l|l|l|l|l|l|l|l|l|l|l|l|l|l|l|l|}
\hline
                         & \multicolumn{2}{c|}{\textbf{Inclusiveness}} & \multicolumn{2}{c|}{\textbf{\begin{tabular}[c]{@{}l@{}}Privacy/Security\end{tabular}}} & \multicolumn{2}{c|}{\textbf{Compatibility}} & \multicolumn{2}{c|}{\textbf{\begin{tabular}[c]{@{}l@{}}Location/Language\end{tabular}}} & \multicolumn{2}{c|}{\textbf{Preference}} & \multicolumn{2}{c|}{\textbf{Satisfaction}} & \multicolumn{2}{c|}{\textbf{\begin{tabular}[c]{@{}l@{}}Emotional  Aspects\end{tabular}}} & \multicolumn{2}{c|}{\textbf{Accessibility}} & \multicolumn{2}{c|}{\textbf{Total}}\\ \hline \hline
\textbf{Project Name}         & \textbf{\#}             & \textbf{\%}            & \textbf{\#}              & \textbf{\%}              & \textbf{\#}             & \textbf{\%}            & \textbf{\#}               & \textbf{\%}              & \textbf{\#}           & \textbf{\%}           & \textbf{\#}            & \textbf{\%}            & \textbf{\#}               & \textbf{\%}              & \textbf{\#}             & \textbf{\%}  & \textbf{\#}             & \textbf{\%}            \\ \hline
\textbf{Firefox}     & 3                       & 0.82                   & 6                        & 1.63                     & 5                       & 1.36                   & 1                         & 0.27                     & 14                    & 3.81                  & 10                     & 2.72                   & 3                         & 0.82                     & 5                       & 1.36      & 47 & 12.80             \\ \hline
\textbf{Cwa-app-ios}     & 14                      & 4.33                   & 11                       & 3.41                     & 13                      & 4.02                   & 18                        & 5.57                     & 5                     & 1.55                  & 17                     & 5.26                   & 3                         & 0.93                     & 4                       & 1.24       & 85 & 26.31              \\ \hline
\textbf{Cwa-app-android} & 3                       & 1.66                   & 21                       & 11.60                    & 10                      & 5.52                   & 19                        & 10.50                    & 6                     & 3.31                  & 11                     & 6.08                   & 1                         & 0.55                     & 1                       & 0.55             & 72 & 39.77        \\ \hline
\textbf{Healthchecks}    & 1                       & 0.34                   & 5                        & 1.71                     & 0                       & 0.00                   & 5                         & 1.71                     & 11                    & 3.75                  & 8                      & 2.73                   & 2                         & 0.68                     & 2                       & 0.68            & 34 & 11.60         \\ \hline
\textbf{COVIDSafe-ios}      & 2                       & 5.88                   & 4                        & 11.76                    & 4                       & 11.76                  & 1                         & 2.94                     & 6                     & 17.65                 & 5                      & 14.71                  & 0                         & 0.00                     & 0                       & 0.00                 & 22 & 64.70    \\ \hline
\textbf{COVIDSafe-android}  & 0                       & 0.00                   & 15                       & 32.61                    & 1                       & 2.17                   & 0                         & 0.00                     & 4                     & 8.70                  & 6                      & 13.04                  & 1                         & 2.17                     & 0                       & 0.00             & 27 & 58.69        \\ \hline
\end{tabular}
}
\end{table*}

\section{Discussion}\label{sec:Discussion}

\textit{\textbf{A wide variety of human-centric issues are discussed in different GitHub repositories;}} 
There are different human-centric issues that developers raise during the development of software for their end-users. 
Some categories, such as inclusiveness, cover a wider range of human factors, for example age, gender, and culture.
~Some issues are more technical-related, such as compatibility, and some are applicable to a more general range of audiences, such as users' satisfaction.
Some of the human-centric issues are discussed more commonly, such as privacy, satisfaction and user preferences, whereas we found there is a limited discussion of accessibility and emotional aspects. Lack of accessibility-related discussions can reflect the fact that disabled people are a small minority of the users \cite{Alshayban:2020}. According to app developers' survey responses in \cite{Alshayban:2020}, accessibility is often not treated as importantly as other aspects of quality, such as security. 
This paper encourages further research for mining and exploring (1) developer-based repositories to understand human-centric discussions and also (2) end-user-based repositories to understand what the diverse end-users actually need. This would help to understand what human-centric issues users ask for and whether they correspond with developer discussions.

\textit{\textbf{Human-centric issues are different across different projects;}}
Our findings show that the prevalence of human-centric issues varies across different projects. Unlike our expectation, projects designed for challenged end-users, e.g., visually impaired and dyslexic users and the ones with large scale end-users, e.g., Firefox, have very limited discussions of human-centric aspects, specifically in accessibility and inclusiveness categories. This encourages future research to study other platforms, both developer-based issues tracking systems, e.g., JIRA and end-user-based app reviews to be able to better understand the needs of the vulnerable end-users.

\textit{\textbf{There is no structured way of reporting and addressing most human-centric issues on GitHub;}}
We found that the human-centric issues are mostly discussed from a technical perspective by the developers. This concern has also been found in the work on usability defect reporting in general \cite{yusop2016reporting}. This indicates the need for a more human-centric issue reporting and follow-up process and tools. Issue reporting systems should include relevant details from not only a technical perspective but also a non-technical end-user understandable point of view. Providing the users with an option to report such issues help the underrepresented groups of users to be engaged and to bring new perspectives on research. Future work is therefore encouraged to incorporate reporting human-centric issues in a systematic way during the software development process. Such reporting tools should, of course, themselves be human-centric and support a diverse range of end-users of the reporting tools.

\textit{\textbf{Awareness of human-centric issues can help developers and researchers to incorporate and report  human-centric issues more effectively;}}
Developers need to be more aware of the human-centric issues of their end-users in order to design more inclusive and human-centred software and to avoid negative impacts on different diverse end-user groups. Software engineers are typically very different from most end-users - a profession heavily dominated by men; relatively young; affluent; technical; most proficient in English; and while some have physical/mental challenges, these are generally different or of less severity than many users, especially for software targeted to vulnerable end-users \cite{grundy2020towards,grundy2020humanise}. These influence the degree that developers appreciate and know how to address human-centric issues to meet the needs of their diverse end-users. Training the developers, supporting them by providing required resources, and increasing their general awareness of the human-centric issues could improve the consideration of these issues during the development process. Results from \cite{Alshayban:2020} indicate the importance of accessibility awareness to make app developers becoming ambassadors of accessibility in their organisations. These will ultimately help to design software that fits the needs of diverse end-users and vulnerable groups of the users in society and promotes better inclusion and belonging.

\section{Threats to Validity}\label{sec:ThreatsValidity}
\textbf{Internal Validity:}  The selection of the 12 studied projects, issue comments from each project, and the qualitative analysis process may have introduced limitations to our study. First, our selection of the projects was motivated to study different types of projects (i.e., they range from small-scale projects to large-scale projects, from projects designed for millions of users to projects targeting vulnerable users). 
Second, manually inspecting all issue discussions was not feasible. Hence, we only analysed a subset of issue discussions from each project. We may have missed important developer human-centric issue discussions or may have found disproportionately many. Third, analysing and labelling the qualitative data might be subjective and error-prone. Our strategy to mitigate this issue was that each issue comment was independently analysed and labelled by two persons. Any disagreements between two analysts on labelling issue comments were resolved either by open discussions or involving the third analyst in the discussions. It should be noted that when it was not clear to identify the type of human-centric issue from a given issue comment, we labelled it as a non-human-centric issue in order to avoid possible risks and mistakes. Hence, we are confident that our classification of human-centric issues is credible with minimum mislabelled issue comments.

\textbf{External Validity:}   Two factors can limit the generalisability of our findings.
Firstly, the 12 selected projects mainly target either large-scale users or vulnerable users. We acknowledge that our findings in this paper may not be generalised to all different types of GitHub projects. For example, developers of software projects to be used by developers, engineers, and scientists (e.g., Jupyter notebook) may consider other aspects of human-centric issues. Secondly, the identified categories of human-centric issues are exclusive to GitHub and are not comprehensive. Hence, analysing other open-source software repositories (e.g., Bitbucket) and software artefacts (e.g., commits, requirement specifications) of proprietary and open-source projects may lead to identifying different and/or a more comprehensive set of human-centric issues categories.

\section{Related Work}\label{sec:RelatedWork}


Online repositories, question and answer sites, and issue tracking platforms such as GitHub, StackOverflow, and Jira not only contain rich data discussing technical aspects of the software development process but also include information that provides insight into the \textbf{social and human aspects} of the software development process \cite{Ortu:2016}. GitHub has been of considerable interest to software engineering researchers for years \cite{kalliamvakou2014promises} due to many open source projects and rich technical and non-technical information that can be mined. Many of the projects hosted on GitHub are public, and therefore anyone can view the activities, including actions around issues, pull requests, and commits within those projects.


Pletea et al. focused on security-related discussions on GitHub, as mined from discussions around commits and pull requests \cite{pletea2014security}. Liao et al. studied username tagging in GitHub projects to understand and model the status and identity of who is speaking and who is being addressed \cite{liao2019status}. Ko et al. analysed developer design discussions through Bugzilla bug reports to understand the design challenges and how the decisions are made to adapt to user needs \cite{ko2011design}. Twidale et al. focused on usability bug reports in Bugzilla \cite{twidale2005exploring} while Andreasen et al. explored developers' opinions about usability through surveys, interviews, and mining software repositories \cite{andreasen2006usability}. Studies have also mined social aspects in repositories. Dabbish et al. mined GitHub for transparency and collaboration in GitHub projects \cite{Dabbish:2012}, while Dam et al. mined open-source projects for social norms \cite{Dam:2015}. Barcellini et al. analysed and visualised social, thematic temporal, and design aspects of online software repositories to understand and model the dynamics of the open source software (OSS) design process in mailing list exchanges \cite{barcellini2008socio}. In their paper, social aspects focus on how roles emerge during discussions, thematic temporal on how themes of discussion emerge, diverge, and are refined over time, and design dynamics on how the online discussions reflect the “workflow” of the project.

Some works have focused on mining and classifying specific human aspects of developers in software repositories and issue tracking platforms. Mining more than 2 million issues in Jira from 4 open-source software projects \cite{Ortu:2015}, Ortu et al. found a positive correlation between developers’ emotions and issue fixing time. Positive emotions resulted in shorter issue-fixing time while negative emotions related to longer issue-fixing time. Cabrera-Diego et al. developed classifiers for comments related to emotions on StackOverflow and Jira. Using features derived from different lexica, their results show significant improvements over the current state of the art in emotion classification \cite{CABRERADIEGO:2020}. Another study analysed software artefacts for the presence of emotional information in the software development process \cite{Murgia:2014}. Results of an analysis of the Apache Software Foundation issue tracking show that developers do express emotion while discussing technical issues. Although these studies focus on a specific human aspect (i.e., emotion) from a developer's perspective, they indicate that a rational view of the software development process is insufficient; human aspects such as emotions can negatively or positively affect the development process and be propagated into the resulting software artefact, e.g., happiness, a positive emotion, increases creativity \cite{Fredrickson2001}, which is good for a successful software design \cite{Brooks:1987}. 


Addressing the role of technical proficiency in the software development process, Rocetti et al. compared two approaches in participatory design of a large software artefact involving: 1) novice users, and 2) expert users. Their results show that most of the innovative proposals came from novice users \cite{Roccetti:2020}. This shows that designing human-centric software artefacts requires a more participation from novice users, in contrast to the traditional opinion that expert users provide more reliable contribution to the software design process. Alshayban et al. conducted a large-scale study to understand the state of accessibility in android apps and found that accessibility issues are rife in the 1,000 apps they studied. In some cases, mobile app developers are not educated in accessibility principles and/or are not incentivised by their organisations to make their apps more accessible \cite{Alshayban:2020}. Similarly, Rauf et al. analysed a dataset of app developers to examine the rationale behind developers’ prioritisation of security in the software development process \cite{Rauf:2020}. The study shows that social considerations, e.g., fear of users, influenced developers’ reasoning in development activities, including security choices \cite{Rauf:2020}.
More recently, a study on the reflection of human values in mobile app reviews shows that a quarter of the 22,119 app reviews analysed contain perceived violation of human values in mobile apps, supporting the recommendation for the use of app reviews as a potential source for mining values requirements in software projects \cite{Obie:2021}.  


All of the studies discussed above focus on different human and social aspects and provide insight into how these aspects are represented in the software development process and repositories. 
However, none of these works provide an analysis of how human-centric aspects of the end-users are discussed by developers in repositories like GitHub. 
In addition, there currently does not exist a taxonomy of human-centric issues on GitHub or other repositories. Our work fills this important gap by providing a broader view perspective of these discussions, with a focus on \textbf{end-user human-centric issues}. While we do not propose this work to be final and immutable in its current form, it is the first to present a taxonomy of human-centric issues on GitHub. In this paper, we developed categories for these human aspects based on a manual analysis of issue comments from different software projects on GitHub and examined how and to what extent human-centric issues are discussed by developers.

\section{Conclusion}\label{sec:Conclusion}
Based on a manual analysis of 1,691 issue comments from 12 different GitHub repositories, we investigated 
what human-centric issues are discussed by developers and reflected on the fact that there is no standard way of reporting and addressing human-centric issues in GitHub repositories. We categorised the human-centric issues discussed by developers in GitHub repositories into eight different categories: inclusiveness, privacy \& security, compatibility, location \& language, preference, satisfaction, emotional aspects, and accessibility. In our future work, we plan to study human-centric issues raised by the end-users of the same projects in the corresponding app reviews and analyse how they are related. We also plan to investigate other repositories, question and answer sites, and issue tracking platforms, such as Jira and Stack Overflow. We want to explore whether developers with very different human aspects to many of their end-users can be better helped to recognise, appreciate and understand how to address these diverse software end-user human-centric issues.


\section*{Acknowledgment}
Support for this work from ARC Laureate Program FL190100035 is gratefully acknowledged.



%

\bibliographystyle{ACM-Reference-Format}
\bibliography{References}


\begin{thebibliography}{46}


\ifx \showCODEN    \undefined \def \showCODEN     #1{\unskip}     \fi
\ifx \showDOI      \undefined \def \showDOI       #1{#1}\fi
\ifx \showISBNx    \undefined \def \showISBNx     #1{\unskip}     \fi
\ifx \showISBNxiii \undefined \def \showISBNxiii  #1{\unskip}     \fi
\ifx \showISSN     \undefined \def \showISSN      #1{\unskip}     \fi
\ifx \showLCCN     \undefined \def \showLCCN      #1{\unskip}     \fi
\ifx \shownote     \undefined \def \shownote      #1{#1}          \fi
\ifx \showarticletitle \undefined \def \showarticletitle #1{#1}   \fi
\ifx \showURL      \undefined \def \showURL       {\relax}        \fi
\providecommand\bibfield[2]{#2}
\providecommand\bibinfo[2]{#2}
\providecommand\natexlab[1]{#1}
\providecommand\showeprint[2][]{arXiv:#2}

\bibitem[\protect\citeauthoryear{Alshayban, Ahmed, and Malek}{Alshayban
  et~al\mbox{.}}{2020}]%
        {Alshayban:2020}
\bibfield{author}{\bibinfo{person}{Abdulaziz Alshayban},
  \bibinfo{person}{Iftekhar Ahmed}, {and} \bibinfo{person}{Sam Malek}.}
  \bibinfo{year}{2020}\natexlab{}.
\newblock \showarticletitle{Accessibility Issues in Android Apps: State of
  Affairs, Sentiments, and Ways Forward}. In
  \bibinfo{booktitle}{\emph{Proceedings of the ACM/IEEE 42nd International
  Conference on Software Engineering}} (Seoul, South Korea)
  \emph{(\bibinfo{series}{ICSE '20})}. \bibinfo{pages}{1323–1334}.
\newblock
\urldef\tempurl%
\url{https://doi.org/10.1145/3377811.3380392}
\showDOI{\tempurl}


\bibitem[\protect\citeauthoryear{Andreasen, Nielsen, Schr{\o}der, and
  Stage}{Andreasen et~al\mbox{.}}{2006}]%
        {andreasen2006usability}
\bibfield{author}{\bibinfo{person}{Morten~Sieker Andreasen},
  \bibinfo{person}{Henrik~Villemann Nielsen}, \bibinfo{person}{Simon~Ormholt
  Schr{\o}der}, {and} \bibinfo{person}{Jan Stage}.}
  \bibinfo{year}{2006}\natexlab{}.
\newblock \showarticletitle{Usability in open source software development:
  opinions and practice}.
\newblock \bibinfo{journal}{\emph{Information technology and control}}
  \bibinfo{volume}{35}, \bibinfo{number}{3} (\bibinfo{year}{2006}).
\newblock


\bibitem[\protect\citeauthoryear{Barcellini, D{\'e}tienne, Burkhardt, and
  Sack}{Barcellini et~al\mbox{.}}{2008}]%
        {barcellini2008socio}
\bibfield{author}{\bibinfo{person}{Flore Barcellini},
  \bibinfo{person}{Fran{\c{c}}oise D{\'e}tienne}, \bibinfo{person}{Jean-Marie
  Burkhardt}, {and} \bibinfo{person}{Warren Sack}.}
  \bibinfo{year}{2008}\natexlab{}.
\newblock \showarticletitle{A socio-cognitive analysis of online design
  discussions in an Open Source Software community}.
\newblock \bibinfo{journal}{\emph{Interacting with computers}}
  \bibinfo{volume}{20}, \bibinfo{number}{1} (\bibinfo{year}{2008}),
  \bibinfo{pages}{141--165}.
\newblock


\bibitem[\protect\citeauthoryear{Brooks}{Brooks}{1987}]%
        {Brooks:1987}
\bibfield{author}{\bibinfo{person}{Frederick~P. Brooks}.}
  \bibinfo{year}{1987}\natexlab{}.
\newblock \showarticletitle{No Silver Bullet Essence and Accidents of Software
  Engineering}.
\newblock \bibinfo{journal}{\emph{Computer}} \bibinfo{volume}{20},
  \bibinfo{number}{4} (\bibinfo{date}{April} \bibinfo{year}{1987}),
  \bibinfo{pages}{10–19}.
\newblock
\showISSN{0018-9162}
\urldef\tempurl%
\url{https://doi.org/10.1109/MC.1987.1663532}
\showDOI{\tempurl}


\bibitem[\protect\citeauthoryear{Brunet, Murphy, Terra, Figueiredo, and
  Serey}{Brunet et~al\mbox{.}}{2014}]%
        {brunet2014developers}
\bibfield{author}{\bibinfo{person}{Jo{\~a}o Brunet}, \bibinfo{person}{Gail~C
  Murphy}, \bibinfo{person}{Ricardo Terra}, \bibinfo{person}{Jorge Figueiredo},
  {and} \bibinfo{person}{Dalton Serey}.} \bibinfo{year}{2014}\natexlab{}.
\newblock \showarticletitle{Do developers discuss design?}. In
  \bibinfo{booktitle}{\emph{Proceedings of the 11th Working Conference on
  Mining Software Repositories}}. \bibinfo{pages}{340--343}.
\newblock


\bibitem[\protect\citeauthoryear{Cabrera-Diego, Bessis, and
  Korkontzelos}{Cabrera-Diego et~al\mbox{.}}{2020}]%
        {CABRERADIEGO:2020}
\bibfield{author}{\bibinfo{person}{Luis~Adrián Cabrera-Diego},
  \bibinfo{person}{Nik Bessis}, {and} \bibinfo{person}{Ioannis Korkontzelos}.}
  \bibinfo{year}{2020}\natexlab{}.
\newblock \showarticletitle{Classifying emotions in Stack Overflow and JIRA
  using a multi-label approach}.
\newblock \bibinfo{journal}{\emph{Knowledge-Based Systems}}
  \bibinfo{volume}{195} (\bibinfo{year}{2020}), \bibinfo{pages}{105633}.
\newblock
\showISSN{0950-7051}
\urldef\tempurl%
\url{https://doi.org/10.1016/j.knosys.2020.105633}
\showDOI{\tempurl}


\bibitem[\protect\citeauthoryear{Cho, Ippolito, and Yu}{Cho
  et~al\mbox{.}}{2020}]%
        {cho2020contact}
\bibfield{author}{\bibinfo{person}{Hyunghoon Cho}, \bibinfo{person}{Daphne
  Ippolito}, {and} \bibinfo{person}{Yun~William Yu}.}
  \bibinfo{year}{2020}\natexlab{}.
\newblock \showarticletitle{Contact tracing mobile apps for COVID-19: Privacy
  considerations and related trade-offs}.
\newblock \bibinfo{journal}{\emph{arXiv preprint arXiv:2003.11511}}
  (\bibinfo{year}{2020}).
\newblock


\bibitem[\protect\citeauthoryear{Curumsing, Fernando, Abdelrazek, Vasa,
  Mouzakis, and Grundy}{Curumsing et~al\mbox{.}}{2019}]%
        {curumsing2019emotion}
\bibfield{author}{\bibinfo{person}{Maheswaree~Kissoon Curumsing},
  \bibinfo{person}{Niroshinie Fernando}, \bibinfo{person}{Mohamed Abdelrazek},
  \bibinfo{person}{Rajesh Vasa}, \bibinfo{person}{Kon Mouzakis}, {and}
  \bibinfo{person}{John Grundy}.} \bibinfo{year}{2019}\natexlab{}.
\newblock \showarticletitle{Emotion-oriented requirements engineering: A case
  study in developing a smart home system for the elderly}.
\newblock \bibinfo{journal}{\emph{Journal of systems and software}}
  \bibinfo{volume}{147} (\bibinfo{year}{2019}), \bibinfo{pages}{215--229}.
\newblock


\bibitem[\protect\citeauthoryear{Dabbish, Stuart, Tsay, and Herbsleb}{Dabbish
  et~al\mbox{.}}{2012}]%
        {Dabbish:2012}
\bibfield{author}{\bibinfo{person}{Laura Dabbish}, \bibinfo{person}{Colleen
  Stuart}, \bibinfo{person}{Jason Tsay}, {and} \bibinfo{person}{Jim Herbsleb}.}
  \bibinfo{year}{2012}\natexlab{}.
\newblock \showarticletitle{Social Coding in GitHub: Transparency and
  Collaboration in an Open Software Repository}. In
  \bibinfo{booktitle}{\emph{Proceedings of the ACM 2012 Conference on Computer
  Supported Cooperative Work}} (Seattle, Washington, USA)
  \emph{(\bibinfo{series}{CSCW '12})}. \bibinfo{pages}{1277–1286}.
\newblock
\urldef\tempurl%
\url{https://doi.org/10.1145/2145204.2145396}
\showDOI{\tempurl}


\bibitem[\protect\citeauthoryear{{Dam}, {Savarimuthu}, {Avery}, and
  {Ghose}}{{Dam} et~al\mbox{.}}{2015}]%
        {Dam:2015}
\bibfield{author}{\bibinfo{person}{H.~K. {Dam}}, \bibinfo{person}{B.~T.~R.
  {Savarimuthu}}, \bibinfo{person}{D. {Avery}}, {and} \bibinfo{person}{A.
  {Ghose}}.} \bibinfo{year}{2015}\natexlab{}.
\newblock \showarticletitle{Mining Software Repositories for Social Norms}. In
  \bibinfo{booktitle}{\emph{2015 IEEE/ACM 37th IEEE International Conference on
  Software Engineering}}, Vol.~\bibinfo{volume}{2}. \bibinfo{pages}{627--630}.
\newblock


\bibitem[\protect\citeauthoryear{Fredrickson}{Fredrickson}{2001}]%
        {Fredrickson2001}
\bibfield{author}{\bibinfo{person}{B. Fredrickson}.}
  \bibinfo{year}{2001}\natexlab{}.
\newblock \showarticletitle{The role of positive emotions in positive
  psychology. The broaden-and-build theory of positive emotions.}
\newblock \bibinfo{journal}{\emph{The American psychologist}}
  \bibinfo{volume}{56}, \bibinfo{number}{3} (\bibinfo{year}{2001}),
  \bibinfo{pages}{218--26}.
\newblock


\bibitem[\protect\citeauthoryear{Glaser, Strauss, and Strutzel}{Glaser
  et~al\mbox{.}}{1968}]%
        {glaser1968discovery}
\bibfield{author}{\bibinfo{person}{Barney~G Glaser}, \bibinfo{person}{Anselm~L
  Strauss}, {and} \bibinfo{person}{Elizabeth Strutzel}.}
  \bibinfo{year}{1968}\natexlab{}.
\newblock \showarticletitle{The discovery of grounded theory; strategies for
  qualitative research}.
\newblock \bibinfo{journal}{\emph{Nursing research}} \bibinfo{volume}{17},
  \bibinfo{number}{4} (\bibinfo{year}{1968}), \bibinfo{pages}{364}.
\newblock


\bibitem[\protect\citeauthoryear{Grundy, Khalajzadeh, and Mcintosh}{Grundy
  et~al\mbox{.}}{2020a}]%
        {grundy2020towards}
\bibfield{author}{\bibinfo{person}{John Grundy}, \bibinfo{person}{Hourieh
  Khalajzadeh}, {and} \bibinfo{person}{Jennifer Mcintosh}.}
  \bibinfo{year}{2020}\natexlab{a}.
\newblock \showarticletitle{Towards Human-centric Model-driven Software
  Engineering.}. In \bibinfo{booktitle}{\emph{ENASE}}.
  \bibinfo{pages}{229--238}.
\newblock


\bibitem[\protect\citeauthoryear{Grundy, Khalajzadeh, McIntosh, Kanij, and
  Mueller}{Grundy et~al\mbox{.}}{2020b}]%
        {grundy2020humanise}
\bibfield{author}{\bibinfo{person}{John Grundy}, \bibinfo{person}{Hourieh
  Khalajzadeh}, \bibinfo{person}{Jennifer McIntosh}, \bibinfo{person}{Tanjila
  Kanij}, {and} \bibinfo{person}{Ingo Mueller}.}
  \bibinfo{year}{2020}\natexlab{b}.
\newblock \showarticletitle{Humanise: Approaches to achieve more human-centric
  software engineering}. In \bibinfo{booktitle}{\emph{International Conference
  on Evaluation of Novel Approaches to Software Engineering}}. Springer,
  \bibinfo{pages}{444--468}.
\newblock


\bibitem[\protect\citeauthoryear{Hartzel}{Hartzel}{2003}]%
        {hartzel2003self}
\bibfield{author}{\bibinfo{person}{Kathleen Hartzel}.}
  \bibinfo{year}{2003}\natexlab{}.
\newblock \showarticletitle{How self-efficacy and gender issues affect software
  adoption and use}.
\newblock \bibinfo{journal}{\emph{Commun. ACM}} \bibinfo{volume}{46},
  \bibinfo{number}{9} (\bibinfo{year}{2003}), \bibinfo{pages}{167--171}.
\newblock


\bibitem[\protect\citeauthoryear{Humbatova, Jahangirova, Bavota, Riccio,
  Stocco, and Tonella}{Humbatova et~al\mbox{.}}{2020}]%
        {humbatova2020taxonomy}
\bibfield{author}{\bibinfo{person}{Nargiz Humbatova}, \bibinfo{person}{Gunel
  Jahangirova}, \bibinfo{person}{Gabriele Bavota}, \bibinfo{person}{Vincenzo
  Riccio}, \bibinfo{person}{Andrea Stocco}, {and} \bibinfo{person}{Paolo
  Tonella}.} \bibinfo{year}{2020}\natexlab{}.
\newblock \showarticletitle{Taxonomy of real faults in deep learning systems}.
  In \bibinfo{booktitle}{\emph{Proceedings of the ACM/IEEE 42nd International
  Conference on Software Engineering}}. \bibinfo{pages}{1110--1121}.
\newblock


\bibitem[\protect\citeauthoryear{Hussain, Abd~Razak, Mkpojiogu, and
  Hamdi}{Hussain et~al\mbox{.}}{2017}]%
        {hussain2017ux}
\bibfield{author}{\bibinfo{person}{Azham Hussain}, \bibinfo{person}{Mohd
  Nur~Faiz Abd~Razak}, \bibinfo{person}{Emmanuel~OC Mkpojiogu}, {and}
  \bibinfo{person}{Mohd Maizan~Fishol Hamdi}.} \bibinfo{year}{2017}\natexlab{}.
\newblock \showarticletitle{UX evaluation of video streaming application with
  teenage users}.
\newblock \bibinfo{journal}{\emph{Journal of Telecommunication, Electronic and
  Computer Engineering (JTEC)}} \bibinfo{volume}{9}, \bibinfo{number}{2-11}
  (\bibinfo{year}{2017}), \bibinfo{pages}{129--131}.
\newblock


\bibitem[\protect\citeauthoryear{Kadam and Bhalerao}{Kadam and
  Bhalerao}{2010}]%
        {kadam2010sample}
\bibfield{author}{\bibinfo{person}{Prashant Kadam} {and}
  \bibinfo{person}{Supriya Bhalerao}.} \bibinfo{year}{2010}\natexlab{}.
\newblock \showarticletitle{Sample size calculation}.
\newblock \bibinfo{journal}{\emph{International journal of Ayurveda research}}
  \bibinfo{volume}{1}, \bibinfo{number}{1} (\bibinfo{year}{2010}),
  \bibinfo{pages}{55}.
\newblock


\bibitem[\protect\citeauthoryear{Kalliamvakou, Gousios, Blincoe, Singer,
  German, and Damian}{Kalliamvakou et~al\mbox{.}}{2014}]%
        {kalliamvakou2014promises}
\bibfield{author}{\bibinfo{person}{Eirini Kalliamvakou},
  \bibinfo{person}{Georgios Gousios}, \bibinfo{person}{Kelly Blincoe},
  \bibinfo{person}{Leif Singer}, \bibinfo{person}{Daniel~M German}, {and}
  \bibinfo{person}{Daniela Damian}.} \bibinfo{year}{2014}\natexlab{}.
\newblock \showarticletitle{The promises and perils of mining GitHub}. In
  \bibinfo{booktitle}{\emph{Proceedings of the 11th working conference on
  mining software repositories}}. \bibinfo{pages}{92--101}.
\newblock


\bibitem[\protect\citeauthoryear{Khalajzadeh, Shahin, Obie, and
  Grundy}{Khalajzadeh et~al\mbox{.}}{2021}]%
        {anonymous_2021_4739069}
\bibfield{author}{\bibinfo{person}{Hourieh Khalajzadeh},
  \bibinfo{person}{Mojtaba Shahin}, \bibinfo{person}{Humphrey Obie}, {and}
  \bibinfo{person}{John Grundy}.} \bibinfo{year}{2021}\natexlab{}.
\newblock \bibinfo{booktitle}{\emph{{What Do Developers Discuss about End-User
  Human-Centric Issues on GitHub?}}}
\newblock
\urldef\tempurl%
\url{https://doi.org/10.5281/zenodo.4739069}
\showDOI{\tempurl}


\bibitem[\protect\citeauthoryear{Khomh, Dhaliwal, Zou, and Adams}{Khomh
  et~al\mbox{.}}{2012}]%
        {khomh2012faster}
\bibfield{author}{\bibinfo{person}{Foutse Khomh}, \bibinfo{person}{Tejinder
  Dhaliwal}, \bibinfo{person}{Ying Zou}, {and} \bibinfo{person}{Bram Adams}.}
  \bibinfo{year}{2012}\natexlab{}.
\newblock \showarticletitle{Do faster releases improve software quality? an
  empirical case study of mozilla firefox}. In \bibinfo{booktitle}{\emph{2012
  9th IEEE Working Conference on Mining Software Repositories (MSR)}}. IEEE,
  \bibinfo{pages}{179--188}.
\newblock


\bibitem[\protect\citeauthoryear{Ko and Chilana}{Ko and Chilana}{2011}]%
        {ko2011design}
\bibfield{author}{\bibinfo{person}{Andrew~J Ko} {and} \bibinfo{person}{Parmit~K
  Chilana}.} \bibinfo{year}{2011}\natexlab{}.
\newblock \showarticletitle{Design, discussion, and dissent in open bug
  reports}.
\newblock In \bibinfo{booktitle}{\emph{Proceedings of the 2011 iConference}}.
  \bibinfo{pages}{106--113}.
\newblock


\bibitem[\protect\citeauthoryear{Kulyk, Kosara, Urquiza, and Wassink}{Kulyk
  et~al\mbox{.}}{2007}]%
        {2007Kulyk}
\bibfield{author}{\bibinfo{person}{Olga Kulyk}, \bibinfo{person}{Robert
  Kosara}, \bibinfo{person}{Jaime Urquiza}, {and} \bibinfo{person}{Ingo
  Wassink}.} \bibinfo{year}{2007}\natexlab{}.
\newblock \showarticletitle{Human-centered aspects}.
\newblock In \bibinfo{booktitle}{\emph{Human-centered visualization
  environments}}. \bibinfo{publisher}{Springer}, \bibinfo{pages}{13--75}.
\newblock


\bibitem[\protect\citeauthoryear{Lewis and Wyatt}{Lewis and Wyatt}{2014}]%
        {lewis2014mhealth}
\bibfield{author}{\bibinfo{person}{Thomas~Lorchan Lewis} {and}
  \bibinfo{person}{Jeremy~C Wyatt}.} \bibinfo{year}{2014}\natexlab{}.
\newblock \showarticletitle{mHealth and mobile medical apps: a framework to
  assess risk and promote safer use}.
\newblock \bibinfo{journal}{\emph{Journal of medical Internet research}}
  \bibinfo{volume}{16}, \bibinfo{number}{9} (\bibinfo{year}{2014}),
  \bibinfo{pages}{e210}.
\newblock


\bibitem[\protect\citeauthoryear{Liao, Yang, Kavaler, Filkov, and Devanbu}{Liao
  et~al\mbox{.}}{2019}]%
        {liao2019status}
\bibfield{author}{\bibinfo{person}{Jingxian Liao}, \bibinfo{person}{Guowei
  Yang}, \bibinfo{person}{David Kavaler}, \bibinfo{person}{Vladimir Filkov},
  {and} \bibinfo{person}{Prem Devanbu}.} \bibinfo{year}{2019}\natexlab{}.
\newblock \showarticletitle{Status, identity, and language: A study of issue
  discussions in GitHub}.
\newblock \bibinfo{journal}{\emph{PloS one}} \bibinfo{volume}{14},
  \bibinfo{number}{6} (\bibinfo{year}{2019}), \bibinfo{pages}{e0215059}.
\newblock


\bibitem[\protect\citeauthoryear{Maguire}{Maguire}{2013}]%
        {maguire2013using}
\bibfield{author}{\bibinfo{person}{Martin Maguire}.}
  \bibinfo{year}{2013}\natexlab{}.
\newblock \showarticletitle{Using human factors standards to support user
  experience and agile design}. In \bibinfo{booktitle}{\emph{International
  Conference on Universal Access in Human-Computer Interaction}}. Springer,
  \bibinfo{pages}{185--194}.
\newblock


\bibitem[\protect\citeauthoryear{Miller, Pedell, Lopez-Lorca, Mendoza,
  Sterling, and Keirnan}{Miller et~al\mbox{.}}{2015}]%
        {miller2015emotion}
\bibfield{author}{\bibinfo{person}{Tim Miller}, \bibinfo{person}{Sonja Pedell},
  \bibinfo{person}{Antonio~A Lopez-Lorca}, \bibinfo{person}{Antonette Mendoza},
  \bibinfo{person}{Leon Sterling}, {and} \bibinfo{person}{Alen Keirnan}.}
  \bibinfo{year}{2015}\natexlab{}.
\newblock \showarticletitle{Emotion-led modelling for people-oriented
  requirements engineering: the case study of emergency systems}.
\newblock \bibinfo{journal}{\emph{Journal of Systems and Software}}
  \bibinfo{volume}{105} (\bibinfo{year}{2015}), \bibinfo{pages}{54--71}.
\newblock


\bibitem[\protect\citeauthoryear{Mo, Shen, Chen, and Zhu}{Mo
  et~al\mbox{.}}{2015}]%
        {mo2015tbil}
\bibfield{author}{\bibinfo{person}{Wenkai Mo}, \bibinfo{person}{Beijun Shen},
  \bibinfo{person}{Yuting Chen}, {and} \bibinfo{person}{Jiangang Zhu}.}
  \bibinfo{year}{2015}\natexlab{}.
\newblock \showarticletitle{Tbil: A tagging-based approach to identity linkage
  across software communities}. In \bibinfo{booktitle}{\emph{2015 Asia-Pacific
  Software Engineering Conference (APSEC)}}. IEEE, \bibinfo{pages}{56--63}.
\newblock


\bibitem[\protect\citeauthoryear{Murgia, Tourani, Adams, and Ortu}{Murgia
  et~al\mbox{.}}{2014}]%
        {Murgia:2014}
\bibfield{author}{\bibinfo{person}{Alessandro Murgia},
  \bibinfo{person}{Parastou Tourani}, \bibinfo{person}{Bram Adams}, {and}
  \bibinfo{person}{Marco Ortu}.} \bibinfo{year}{2014}\natexlab{}.
\newblock \showarticletitle{Do Developers Feel Emotions? An Exploratory
  Analysis of Emotions in Software Artifacts}. In
  \bibinfo{booktitle}{\emph{Proceedings of the 11th Working Conference on
  Mining Software Repositories}} (Hyderabad, India) \emph{(\bibinfo{series}{MSR
  2014})}. \bibinfo{pages}{262–271}.
\newblock
\showISBNx{9781450328630}
\urldef\tempurl%
\url{https://doi.org/10.1145/2597073.2597086}
\showDOI{\tempurl}


\bibitem[\protect\citeauthoryear{Obie, Hussain, Xia, Grundy, Li, Turhan,
  Whittle, and Shahin}{Obie et~al\mbox{.}}{2021}]%
        {Obie:2021}
\bibfield{author}{\bibinfo{person}{Humphrey~O. Obie}, \bibinfo{person}{Waqar
  Hussain}, \bibinfo{person}{Xin Xia}, \bibinfo{person}{John Grundy},
  \bibinfo{person}{Li Li}, \bibinfo{person}{Burak Turhan}, \bibinfo{person}{Jon
  Whittle}, {and} \bibinfo{person}{Mojtaba Shahin}.}
  \bibinfo{year}{2021}\natexlab{}.
\newblock \showarticletitle{A First Look at Human Values-Violation in App
  Reviews}. In \bibinfo{booktitle}{\emph{ICSE-SEIS}}.
\newblock


\bibitem[\protect\citeauthoryear{Ortu, Destefanis, Adams, Murgia, Marchesi, and
  Tonelli}{Ortu et~al\mbox{.}}{2015}]%
        {Ortu:2015}
\bibfield{author}{\bibinfo{person}{Marco Ortu}, \bibinfo{person}{Giuseppe
  Destefanis}, \bibinfo{person}{Bram Adams}, \bibinfo{person}{Alessandro
  Murgia}, \bibinfo{person}{Michele Marchesi}, {and} \bibinfo{person}{Roberto
  Tonelli}.} \bibinfo{year}{2015}\natexlab{}.
\newblock \showarticletitle{The JIRA Repository Dataset: Understanding Social
  Aspects of Software Development}. In \bibinfo{booktitle}{\emph{Proceedings of
  the 11th International Conference on Predictive Models and Data Analytics in
  Software Engineering}} (Beijing, China) \emph{(\bibinfo{series}{PROMISE
  '15})}. Article \bibinfo{articleno}{1}, \bibinfo{numpages}{4}~pages.
\newblock
\showISBNx{9781450337151}
\urldef\tempurl%
\url{https://doi.org/10.1145/2810146.2810147}
\showDOI{\tempurl}


\bibitem[\protect\citeauthoryear{Ortu, Murgia, Destefanis, Tourani, Tonelli,
  Marchesi, and Adams}{Ortu et~al\mbox{.}}{2016}]%
        {Ortu:2016}
\bibfield{author}{\bibinfo{person}{Marco Ortu}, \bibinfo{person}{Alessandro
  Murgia}, \bibinfo{person}{Giuseppe Destefanis}, \bibinfo{person}{Parastou
  Tourani}, \bibinfo{person}{Roberto Tonelli}, \bibinfo{person}{Michele
  Marchesi}, {and} \bibinfo{person}{Bram Adams}.}
  \bibinfo{year}{2016}\natexlab{}.
\newblock \showarticletitle{The Emotional Side of Software Developers in JIRA}.
  In \bibinfo{booktitle}{\emph{Proceedings of the 13th International Conference
  on Mining Software Repositories}} (Austin, Texas) \emph{(\bibinfo{series}{MSR
  '16})}. \bibinfo{pages}{480–483}.
\newblock
\showISBNx{9781450341868}
\urldef\tempurl%
\url{https://doi.org/10.1145/2901739.2903505}
\showDOI{\tempurl}


\bibitem[\protect\citeauthoryear{{\O}vad, Bornoe, Larsen, and Stage}{{\O}vad
  et~al\mbox{.}}{2015}]%
        {ovad2015teaching}
\bibfield{author}{\bibinfo{person}{Tina {\O}vad}, \bibinfo{person}{Nis Bornoe},
  \bibinfo{person}{Lars~Bo Larsen}, {and} \bibinfo{person}{Jan Stage}.}
  \bibinfo{year}{2015}\natexlab{}.
\newblock \showarticletitle{Teaching software developers to perform UX tasks}.
  In \bibinfo{booktitle}{\emph{Proceedings of the Annual Meeting of the
  Australian Special Interest Group for Computer Human Interaction}}.
  \bibinfo{pages}{397--406}.
\newblock


\bibitem[\protect\citeauthoryear{Parker, Fraser, Abeler-D{\"o}rner, and
  Bonsall}{Parker et~al\mbox{.}}{2020}]%
        {parker2020ethics}
\bibfield{author}{\bibinfo{person}{Michael~J Parker},
  \bibinfo{person}{Christophe Fraser}, \bibinfo{person}{Lucie
  Abeler-D{\"o}rner}, {and} \bibinfo{person}{David Bonsall}.}
  \bibinfo{year}{2020}\natexlab{}.
\newblock \showarticletitle{Ethics of instantaneous contact tracing using
  mobile phone apps in the control of the COVID-19 pandemic}.
\newblock \bibinfo{journal}{\emph{Journal of Medical Ethics}}
  \bibinfo{volume}{46}, \bibinfo{number}{7} (\bibinfo{year}{2020}),
  \bibinfo{pages}{427--431}.
\newblock


\bibitem[\protect\citeauthoryear{Perez}{Perez}{2019}]%
        {perez2019invisible}
\bibfield{author}{\bibinfo{person}{Caroline~Criado Perez}.}
  \bibinfo{year}{2019}\natexlab{}.
\newblock \bibinfo{booktitle}{\emph{Invisible women: Exposing data bias in a
  world designed for men}}.
\newblock \bibinfo{publisher}{Random House}.
\newblock


\bibitem[\protect\citeauthoryear{Pletea, Vasilescu, and Serebrenik}{Pletea
  et~al\mbox{.}}{2014}]%
        {pletea2014security}
\bibfield{author}{\bibinfo{person}{Daniel Pletea}, \bibinfo{person}{Bogdan
  Vasilescu}, {and} \bibinfo{person}{Alexander Serebrenik}.}
  \bibinfo{year}{2014}\natexlab{}.
\newblock \showarticletitle{Security and emotion: sentiment analysis of
  security discussions on github}. In \bibinfo{booktitle}{\emph{Proceedings of
  the 11th working conference on mining software repositories}}.
  \bibinfo{pages}{348--351}.
\newblock


\bibitem[\protect\citeauthoryear{Prikladnicki, Dittrich, Sharp, De~Souza,
  Cataldo, and Hoda}{Prikladnicki et~al\mbox{.}}{2013}]%
        {rafael2013cooperative}
\bibfield{author}{\bibinfo{person}{Rafael Prikladnicki},
  \bibinfo{person}{Yvonne Dittrich}, \bibinfo{person}{Helen Sharp},
  \bibinfo{person}{Cleidson De~Souza}, \bibinfo{person}{Marcelo Cataldo}, {and}
  \bibinfo{person}{Rashina Hoda}.} \bibinfo{year}{2013}\natexlab{}.
\newblock \showarticletitle{Cooperative and Human Aspects of Software
  Engineering: CHASE 2013}.
\newblock \bibinfo{journal}{\emph{SIGSOFT Softw. Eng. Notes}}
  \bibinfo{volume}{38}, \bibinfo{number}{5} (\bibinfo{date}{Aug.}
  \bibinfo{year}{2013}), \bibinfo{pages}{34–37}.
\newblock
\showISSN{0163-5948}
\urldef\tempurl%
\url{https://doi.org/10.1145/2507288.2507321}
\showDOI{\tempurl}


\bibitem[\protect\citeauthoryear{Rauf, van~der Linden, Levine, Towse, Nuseibeh,
  and Rashid}{Rauf et~al\mbox{.}}{2020}]%
        {Rauf:2020}
\bibfield{author}{\bibinfo{person}{Irum Rauf}, \bibinfo{person}{Dirk van~der
  Linden}, \bibinfo{person}{Mark Levine}, \bibinfo{person}{John Towse},
  \bibinfo{person}{Bashar Nuseibeh}, {and} \bibinfo{person}{Awais Rashid}.}
  \bibinfo{year}{2020}\natexlab{}.
\newblock \showarticletitle{Security but Not for Security's Sake: The Impact of
  Social Considerations on App Developers' Choices}. In
  \bibinfo{booktitle}{\emph{Proceedings of the IEEE/ACM 42nd International
  Conference on Software Engineering Workshops}}
  \emph{(\bibinfo{series}{ICSEW'20})}. \bibinfo{pages}{141–144}.
\newblock


\bibitem[\protect\citeauthoryear{Roccetti, Prandi, Mirri, and
  Salomoni}{Roccetti et~al\mbox{.}}{2020}]%
        {Roccetti:2020}
\bibfield{author}{\bibinfo{person}{M. Roccetti}, \bibinfo{person}{C. Prandi},
  \bibinfo{person}{S. Mirri}, {and} \bibinfo{person}{P. Salomoni}.}
  \bibinfo{year}{2020}\natexlab{}.
\newblock \showarticletitle{Designing human-centric software artifacts with
  future users: a case study}.
\newblock \bibinfo{journal}{\emph{Human-centric Computing and Information
  Sciences}}  \bibinfo{volume}{10} (\bibinfo{year}{2020}),
  \bibinfo{pages}{1--17}.
\newblock


\bibitem[\protect\citeauthoryear{Stock, Davies, Wehmeyer, and Palmer}{Stock
  et~al\mbox{.}}{2008}]%
        {stock2008evaluation}
\bibfield{author}{\bibinfo{person}{Steven~E Stock}, \bibinfo{person}{Daniel~K
  Davies}, \bibinfo{person}{Michael~L Wehmeyer}, {and} \bibinfo{person}{Susan~B
  Palmer}.} \bibinfo{year}{2008}\natexlab{}.
\newblock \showarticletitle{Evaluation of cognitively accessible software to
  increase independent access to cellphone technology for people with
  intellectual disability}.
\newblock \bibinfo{journal}{\emph{Journal of Intellectual Disability Research}}
  \bibinfo{volume}{52}, \bibinfo{number}{12} (\bibinfo{year}{2008}),
  \bibinfo{pages}{1155--1164}.
\newblock


\bibitem[\protect\citeauthoryear{Strengers and Kennedy}{Strengers and
  Kennedy}{2020}]%
        {strengers2020smart}
\bibfield{author}{\bibinfo{person}{Yolande Strengers} {and}
  \bibinfo{person}{Jenny Kennedy}.} \bibinfo{year}{2020}\natexlab{}.
\newblock \bibinfo{booktitle}{\emph{The Smart Wife: Why Siri, Alexa, and Other
  Smart Home Devices Need a Feminist Reboot}}.
\newblock \bibinfo{publisher}{MIT Press}.
\newblock


\bibitem[\protect\citeauthoryear{Tsay, Dabbish, and Herbsleb}{Tsay
  et~al\mbox{.}}{2014}]%
        {tsay2014let}
\bibfield{author}{\bibinfo{person}{Jason Tsay}, \bibinfo{person}{Laura
  Dabbish}, {and} \bibinfo{person}{James Herbsleb}.}
  \bibinfo{year}{2014}\natexlab{}.
\newblock \showarticletitle{Let's talk about it: evaluating contributions
  through discussion in GitHub}. In \bibinfo{booktitle}{\emph{Proceedings of
  the 22nd ACM SIGSOFT international symposium on foundations of software
  engineering}}. \bibinfo{pages}{144--154}.
\newblock


\bibitem[\protect\citeauthoryear{Twidale and Nichols}{Twidale and
  Nichols}{2005}]%
        {twidale2005exploring}
\bibfield{author}{\bibinfo{person}{Michael~B Twidale} {and}
  \bibinfo{person}{David~M Nichols}.} \bibinfo{year}{2005}\natexlab{}.
\newblock \showarticletitle{Exploring usability discussions in open source
  development}. In \bibinfo{booktitle}{\emph{Proceedings of the 38th Annual
  Hawaii International Conference on System Sciences}}. IEEE,
  \bibinfo{pages}{198c--198c}.
\newblock


\bibitem[\protect\citeauthoryear{Wirtz, Jakobs, and Ziefle}{Wirtz
  et~al\mbox{.}}{2009}]%
        {wirtz2009age}
\bibfield{author}{\bibinfo{person}{Simone Wirtz}, \bibinfo{person}{Eva-Maria
  Jakobs}, {and} \bibinfo{person}{Martina Ziefle}.}
  \bibinfo{year}{2009}\natexlab{}.
\newblock \showarticletitle{Age-specific usability issues of software
  interfaces}. In \bibinfo{booktitle}{\emph{Proceedings of the IEA}},
  Vol.~\bibinfo{volume}{17}.
\newblock


\bibitem[\protect\citeauthoryear{Yusop, Grundy, and Vasa}{Yusop
  et~al\mbox{.}}{2016}]%
        {yusop2016reporting}
\bibfield{author}{\bibinfo{person}{Nor Shahida~Mohamad Yusop},
  \bibinfo{person}{John Grundy}, {and} \bibinfo{person}{Rajesh Vasa}.}
  \bibinfo{year}{2016}\natexlab{}.
\newblock \showarticletitle{Reporting usability defects: A systematic
  literature review}.
\newblock \bibinfo{journal}{\emph{IEEE Transactions on Software Engineering}}
  \bibinfo{volume}{43}, \bibinfo{number}{9} (\bibinfo{year}{2016}),
  \bibinfo{pages}{848--867}.
\newblock


\bibitem[\protect\citeauthoryear{Zaman, Adams, and Hassan}{Zaman
  et~al\mbox{.}}{2011}]%
        {zaman2011security}
\bibfield{author}{\bibinfo{person}{Shahed Zaman}, \bibinfo{person}{Bram Adams},
  {and} \bibinfo{person}{Ahmed~E Hassan}.} \bibinfo{year}{2011}\natexlab{}.
\newblock \showarticletitle{Security versus performance bugs: a case study on
  firefox}. In \bibinfo{booktitle}{\emph{Proceedings of the 8th working
  conference on mining software repositories}}. \bibinfo{pages}{93--102}.
\newblock


\end{thebibliography}

\end{document}